\documentclass[modern]{aastex631}

\usepackage{graphicx}

\begin{document}

\title{Prospects for the Crossing of Comet 3I/ATLAS's Ion Tail}

\correspondingauthor{Samuel R. Grant}
\email{samuel.grant@fmi.fi}

\author[0000-0002-3499-8142]{Samuel R. Grant}
\affiliation{Finnish Meteorological Institute\\Erik Palménin aukio 1, 00560 Helsinki, Finland}

\author[0000-0002-5859-1136]{Geraint H. Jones}
\affiliation{European Space Agency, European Space Research and Technology Centre\\ Keplerlaan 1, 2201 AZ Noordwijk, The Netherlands}

\begin{abstract}
During October - November 2025, interstellar comet 3I/ATLAS, will pass upstream of the Europa Clipper and Hera spacecraft. Here, we identify two potential opportunities for in-situ observations of 3I's ion tail by immersion, facilitated by the close alignment between the comet's hyperbolic trajectory with the ecliptic plane. 
During the period 30 October - 6 November 2025, it is predicted that Europa Clipper will potentially be immersed within the ion tail of 3I/ATLAS, providing the opportunity to detect the signatures of an interstellar comet's ion tail. Characteristic changes to the solar wind are also expected to be observed; a magnetic draping structure associated with the comet may be identifiable.  It is further predicted that spacecraft Hera will possibly be immersed within the ion tail of 3I/ATLAS during the period 25 October - 1 November 2025. 
\end{abstract}


\section*{}

Interstellar objects are rare visitors to the Solar System, with only three observed within the last decade. The first, 1I/'Oumuamua,  had no visible coma or outgassing throughout its passage \citep{bannister2019natural}. In contrast, 2I/Borisov had formed a distinct blue coma and ion tail as it reached its perihelion on 8 December 2019 \citep{xing2020water}. 

Comet 3I/ATLAS (C/2025 N1 (ATLAS)) is the third known interstellar object to transit the Solar System \citep{seligman2025ApJ...989L..36S}. Discovered on 1 July 2025, 3I/ATLAS will reach a perihelion distance of 1.36 au around noon UTC on 29 October 2025. Following a retrograde hyperbolic trajectory near the ecliptic, 3I/ATLAS has been observed to have an extensive coma increasing in luminosity \citep{Infooirlabedu,eubanks2025cosmicnoon}. Initial polarimetric observations show the comet is unique and distinct from local comets \citep{gray2025extreme}. 

Opportunities for observations during 3I/ATLAS's transit of the Solar System are collated by \citet{eubanks2025cosmicnoon}. Conditions for observations from near Earth are poor, but as noted by \citet{eubanks2025cosmicnoon}, several spacecraft elsewhere could observe it. Here, we build on Eubank and co-authors' work by providing quantitative information on the possibility of spacecraft crossing for the first time the tail of an interstellar comet. 

Serendipitous ion tail crossings are uncommon events, although several have been catalogued, e.g.  \citet{jones2000, gloeckler2004, matteini2021, JONES2022115199}, including likely crossings by Earth \citep{russell1988geomagnetic,grant2024RNAAS...8..252G}. Such encounters occur when a comet crosses the Sun-spacecraft line on a timeframe such that the solar wind (SW) may transport ions at several hundred km s$^{-1}$ from the comet's coma to a spacecraft. It is likely that many ion tail crossings have passed unnoticed, as identification based solely on in-situ data is difficult. Without context, ion tails' signatures can appear as minor fluctuations in the ambient SW and Heliospheric Magnetic Field. When ions from a comet are detected directly, then the identification is clearer, e.g. \citet{neugebauer07}.

The Tailcatcher program predicts and locates cometary and planetary ion tail encounters. The degree of alignment is quantified by the calculation of the ``impact parameter''. For past crossings, SW velocities downstream of a comet are used to interpolate the SW flow from the spacecraft back towards the Sun. For predictions, typical SW velocities are assumed. The impact parameter is defined by the minimum miss distance from the calculated flow stream to the comet position, and is minimised at closest alignment. Tailcatcher was successfully tested with all known ion tail crossings, and successfully predicted three ion tail encounters \citep{jones2020RNAAS...4...62J, grant2022prediction, grant2024RNAAS...8..252G}. 
\begin{figure}[h!]
\centering
\gridline{\fig{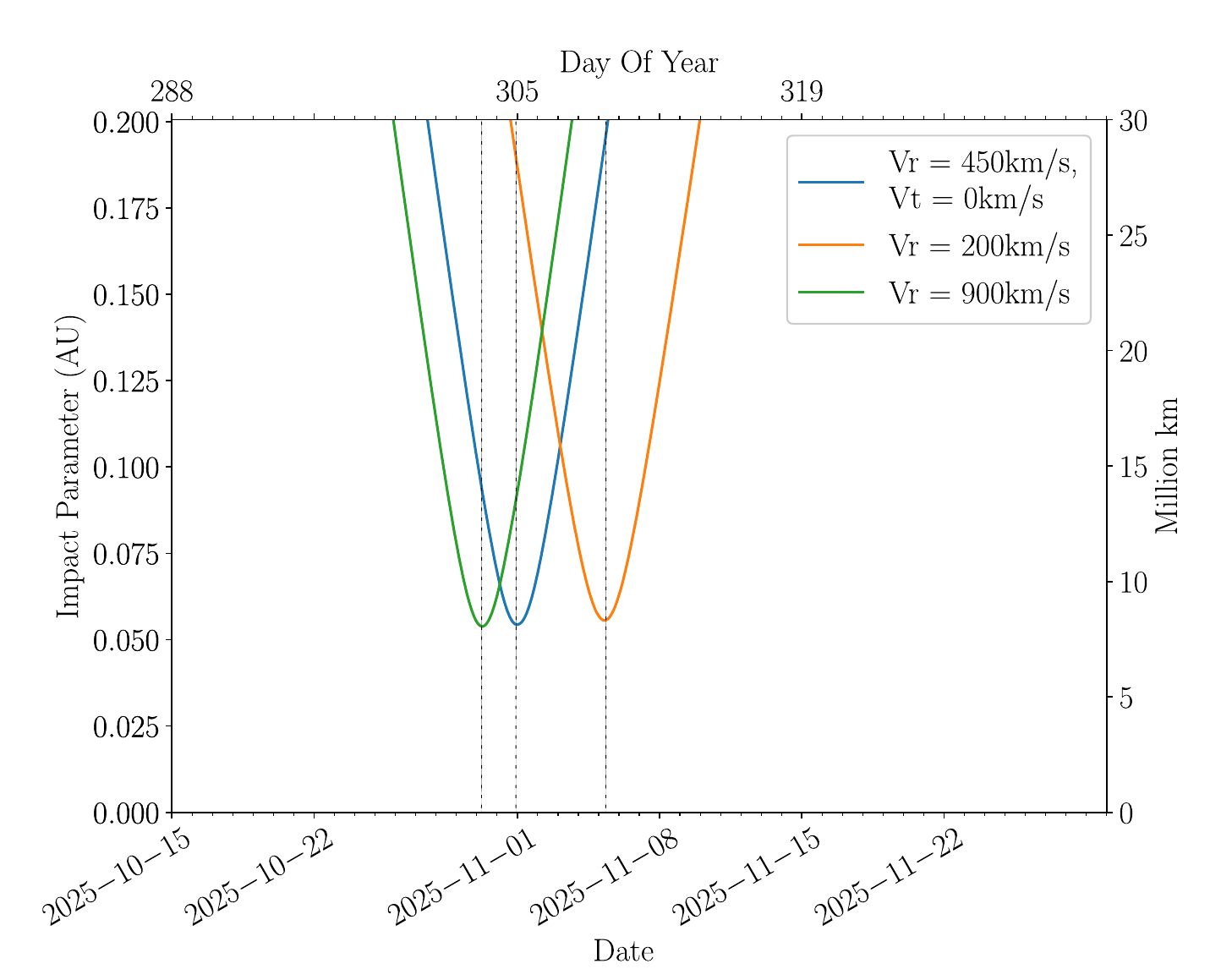}{0.7\textwidth}{}}
\gridline{\fig{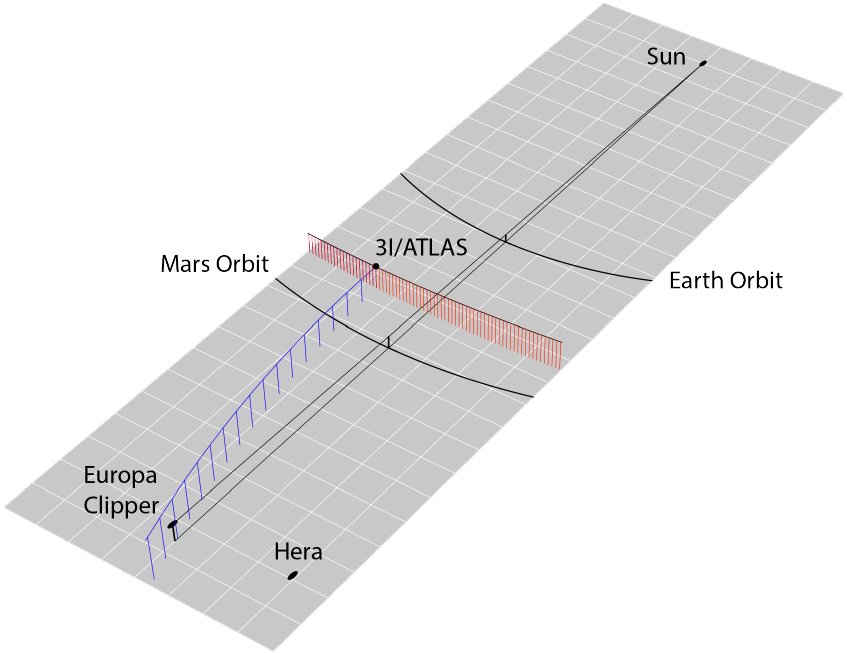}{0.7\textwidth}{}}
    \caption{\emph{(Upper)} Impact parameters for Europa Clipper relative to comet 3I/ATLAS for purely radial SW velocities 200, 450 and 900 km/s. The minima for the three curves fall between 30 October and 6 November 2025. \emph{(Lower)} Simulation of 3I/ATLAS's ion tail on 1 November 2025 assuming a 460~kms$^{-1}$ radial flow. Positions of Europa Clipper, Sun, Hera, and 3I/ATLAS are shown in a frame where the Sun-Europa Clipper line is fixed. The plane represents that of the ecliptic, with grid lines spaced at 0.1~au. The comet's position is shown every 6 hours, with legs showing its position north of the ecliptic. Europa Clipper is also north of the ecliptic by 0.069~au. Earth and Mars's mean orbital distances are included for scale.} \label{combined}
\end{figure}
Fig. \ref{combined} displays miss distances from comet 3I/ATLAS calculated via Tailcatcher for 3 radial SW speeds for Europa Clipper \citep{Clipper2025SSRv..221...25K}. Assuming a perfectly radial flow, higher SW speeds result in earlier and smaller predicted minimum distances from the ion tail axis. In reality, the SW often has significant non-radial components. This can lead to closer or further minimum impact parameters relative to the tail core; this can be assessed after the crossing with measured SW flow vectors. Note that calculations presented here are for the ion tail's main axis; as the tails of active comets can be wide, an encounter with a tail flank of is possible even if its axis is not crossed. The neutral hydrogen corona that surrounds a comet can measure several million km across, and in some cases, tens of millions of km \citep{JONES2022115199}.
The minimum miss distance for SW 'packets' is $\sim$(8.0)$\times$10$^{6}$~km for the dates 30 October -- 6 November 2025, assuming purely radial flow. 
From 18 - 20 August 2025, when 3I/ATLAS was at a heliocentric distance of 2.9 au, the water production rate estimated by \citet{xing2025water} was $(1.36\pm0.35)\times 10^{27} s^{-1}$, quite high for a comet then beyond 2.5 au. Should 3I continue to increase in activity as it approaches perihelion, a broad ion tail is possible, making detection more likely. Spacecraft Hera \citep{Hera2025SSRv..221...70M} is similarly well-aligned with 3I/ATLAS, with its impact parameter reaching a minimum of $\sim$(8.2)$\times$10$^{6}$~km during 25 October -- 1 November 2025.
We note that these calculations are for the comet's ion tail only. No operating spacecraft are expected to be in locations where they could detect dust from 3I.

The encounters described here may be the first direct immersion in material from an interstellar object. Even if pick-up ion measurement is not possible, magnetic field draping and shock signatures can reveal the presence and structure of an ion tail. Europa Clipper carries instruments for plasma \citep{westlake2023SSRv..219...62W} and a magnetometer \citep{kivelson2023SSRv..219...48K}, but Hera is not equipped to detect ions nor the HMF \citep{eubanks2025cosmicnoon}. With favourable SW conditions, Europa Clipper has a rare opportunity to sample an interstellar object's tail. 

It is likely that knowledge of such encounters will become more common as the discovery rate of these objects improves. If a similar object is discovered with a favourable trajectory, the object itself could be encountered at closer range by ESA's Comet Interceptor spacecraft \citep{comet-i2024SSRv..220....9J}.

\begin{acknowledgments}
\section*{Acknowledgments}
Author SRG acknowledges the support of a UKRI STFC research studentship for the development of the Tailcatcher method, and partial support from the European Research Council (Grant agreement No. 101124960). Ephemeris data were obtained via the JPL Horizons system at https://ssd.jpl.nasa.gov/?horizons ; \citet{giorgini1997}.
\end{acknowledgments}

\bibliography{rnaas}{}
\bibliographystyle{aasjournal}

\end{document}